\documentclass[a4paper,10pt,twoside=false]{scrartcl}

\usepackage[utf8]{inputenc}
\usepackage[T1]{fontenc}
\usepackage[kerning=true,protrusion=true,expansion=true]{microtype}

\usepackage[margin=1in]{geometry}
\usepackage{abstract}

\usepackage{amssymb,amsmath,amsthm,yfonts,pifont,bbm,bm,mathrsfs,lettrine,booktabs,siunitx,dsfont}

\usepackage{graphicx,color,framed,mdframed}
\setcapwidth[c]{\textwidth}
\setcapindent{0pt}
\usepackage[dvipsnames]{xcolor}
\graphicspath{{./figs/}{./tikz/}}

\usepackage{soul}

\usepackage{xfp}
\newcommand\SupplementaryMaterials{%
    \renewcommand{\figurename}{Supplementary Figure}
	\xdef\presupequations{\arabic{equation}}
	\xdef\presupfigures{\arabic{figure}}
	\xdef\presupsections{\arabic{section}}
	\renewcommand\theequation{S\fpeval{\arabic{equation}-\presupequations}}
	\renewcommand\thefigure{S\fpeval{\arabic{figure}-\presupfigures}}
	\renewcommand\thesection{Supplementary Note \fpeval{\arabic{section}-\presupsections}}
	\setcounter{page}{1}
	\renewcommand{\thepage}{\arabic{page}}
}
\usepackage[backend=biber,style=phys,articletitle=true,biblabel=brackets,pageranges=true,arxiv=none,mincitenames=1,maxcitenames=1,maxnames=5,uniquename=false,uniquelist=false]{biblatex}
\usepackage[pdftex,
	colorlinks,
	allcolors=blue,
	pdfauthor={A. Rakhubovsky},
	pdftitle={},
	pdfsubject={},
	pdfkeywords={},
	pdfproducer={Latex with hyperref},
	pdfcreator={pdflatex}]{hyperref}

\usepackage[capitalize,poorman]{cleveref}

\newcommand*{\ket}[1]{\left|#1\right\rangle}
\newcommand*{\dd}[1]{\ensuremath{\mathrm{d}#1\:}}
\newcommand*{\da}{\dagger}
\newcommand*{\comm}[2]{\ensuremath{\left[ #1 , #2 \right] }}
\newcommand*{\s}[1]{\ensuremath{_\mathsf{#1}}}
\newcommand*{\up}[1]{\ensuremath{^\text{#1}}}
\newcommand*{\avg}[1]{\left\langle #1 \right\rangle}
\newcommand*{\bra}[1]{\left\langle#1\right|}
\newcommand*{\mel}[3]{\left\langle #1 \middle| #2 \middle| #3 \right\rangle}

\newcommand*{\vk}{\lambda}
\newcommand{\damp}{\eta\s{m}}

\newcommand{\nlgain}{\gamma}

\newcommand{\oper}[1]{\hat{#1}}
\newcommand*{\ox}{\oper x}
\newcommand*{\op}{\oper p}
\newcommand*{\un}{\oper{U}}

\newcommand*{\pmt}[2]{\begin{pmatrix} #1 \\ #2 \end{pmatrix}}

\DeclareMathOperator*{\Ai}{Ai}
\DeclareMathOperator*{\Tr}{\operatorname{Tr}}
\DeclareMathOperator*{\Var}{\operatorname{Var}}

\newcommand{\II}{\mathds{1}}
\newcommand{\oI}{\hat{\II}}

\newcommand{\ee}{\mathrm{e}}
\newcommand{\ii}{\mathrm{i}}

\newenvironment{DIFnomarkup}{}{}

\begin{document}



\setkomafont{author}{\sffamily}

\newcommand{\UPOL}{Department of Optics, Palack{\'y} University, 17. Listopadu 12, 771 46 Olomouc, Czech Republic}
\newcommand{\thetitle}{Stroboscopic high-order nonlinearity for quantum optomechanics}
\title{\thetitle}
\author{
    Andrey A. {Rakhubovsky}
    \footnote{Corresponding author: \href{mailto:andrey.rakhubovsky@gmail.com}{andrey.rakhubovsky@gmail.com}}{ }$^{1}$ \and Radim {Filip}\footnote{\href{mailto:filip@optics.upol.cz}{filip@optics.upol.cz}}{ }$^{1}$
\\
$^1$\emph{Department of Optics, Palack{\'y} University, 17. Listopadu 12, 771 46 Olomouc, Czech Republic}}

\date{}

\renewcommand{\abstractname}{}

\maketitle

\begin{onecolabstract}
	High-order quantum nonlinearity is an important prerequisite for the advanced quantum technology leading to universal quantum processing with large information capacity of continuous variables.
    Levitated optomechanics, a field where motion of dielectric particles is driven by precisely controlled tweezer beams, is capable of attaining the required nonlinearity via engineered potential landscapes of mechanical motion.
	Importantly, to achieve nonlinear quantum effects, the evolution caused by the free motion of mechanics and thermal decoherence have to be suppressed.
	For this purpose, we devise a method of stroboscopic application of a highly nonlinear potential to a mechanical oscillator that leads to the motional quantum non-Gaussian states exhibiting nonclassical negative Wigner function and squeezing of a nonlinear combination of mechanical quadratures.
	We test the method numerically by analysing highly instable cubic potential with relevant experimental parameters of the levitated optomechanics, prove its feasibility within reach, and propose an experimental test.
	The method paves a road for experiments instantaneously transforming a ground state of mechanical oscillators to applicable nonclassical states by nonlinear optical force.
\end{onecolabstract}


\section{Introduction}
Quantum physics and technology with continuous variables (CVs)~\cite{braunsteinquantum2005} has achieved noticeable progress recently.
A potential advantage of CVs is the in principle unlimited energy and information capacity of single oscillator mode.
In order to fully gain the benefits of CVs and to achieve \emph{universal} quantum processing one requires an access to a nonlinear operation~\cite{lloydquantum1999,gottesmanencoding2001}, that is, at least a cubic potential.
Additionally, the CV quantum information processing can be greatly simplified and stabilized if variable higher-order potentials are available~\cite{sefihow2011}.
Variability of nonlinear gates can also help to overcome limits for fault tolerance~\cite{hastrupunsuitability2021}.
Nanomechanical systems profit from a straightforward feasible way to achieve the nonlinearity by inducing a controllable classical nonlinear force of electromagnetic nature on a linear mechanical oscillator~ \cite{meyerresolvedsideband2019,goldwaterlevitated2019,vinanteultralow2020}.
Such a nonlinear force needs to be fast, strong and controllable on demand to access different nonlinearities required for an efficient universal CV quantum processing.
Therefore, the field of optomechanics~\cite{aspelmeyercavity2014,aspelmeyercavity2014a,bowenquantum2015,khaliliquantum2016} is a promising candidate to provide the key element for the variable \emph{on-demand} nonlinearity.
Optomechanical systems have reached a truly quantum domain recently, demonstrating the effects ranging from the ground state cooling~\cite{teufelsideband2011,chanlaser2011} and squeezing~\cite{wollmanquantum2015,pirkkalainensqueezing2015} of the mechanical motion to the entanglement of distant mechanical oscillators~\cite{riedingerremote2018,ockeloen-korppistabilized2018}.
Of particular interest are the levitated systems in which the potential landscape of the mechanical motion is provided by a highly developed device~--- an optical tweezer~\cite{barkercavity2010,changcavity2010,romero-isartoptically2011,millen_quantum_2020}.
Levitated systems have proved useful in force sensing~\cite{mooresearch2014, ranjitzeptonewton2016}, studies of quantum thermodynamics~\cite{gieselerthermal2013,riccioptically2017,gieselerlevitated2018}, testing fundamental physics~\cite{romero-isartquantum2011,batemannearfield2014,goldwatertesting2016} and probing quantum gravity~\cite{bosespin2017,marlettogravitationally2017}.
From the technical point of view, the levitated systems have recently demonstrated noticeable progress in the controllability and engineering, particularly, cooling towards~\cite{jaindirect2016,vovroshparametric2017,windeycavitybased2019,deliccavity2019} and eventually reaching the motional ground state~\cite{deliccooling2020}.
Further theoretical studies of preparation of entangled states of levitated nanoparticles are underway~\cite{rudolphentangling2020,rakhubovskydetecting2020}.
Besides the inherently nonlinear optomechanical interaction met in the standard bulk optomechanical systems the levitated ones possess the attractive possibility of engineering the nonlinear trapping potential~\cite{berutexperimental2012,ryabovthermally2016,ornigottibrownian2018,silerthermally2017,riccioptically2017,silerdiffusing2018,deliccooling2020}.
Moreover, the trapping potentials can be made time-dependent and manipulated at rates exceeding the rate of mechanical decoherence and even the mechanical frequency~\cite{konopik_nonequilibrium_2020}.
In this manuscript, we assume a similar possibility to generate the non-linear potential for a mechanical motion and control it in a fast way (faster than the mechanical frequency).
Our findings do not rely on the specific method of how the nonlinearity is created.

Here we propose a broadly applicable nonlinear stroboscopic method to achieve high-order nonlinearity in optomechanical systems with time-~and space-variable external force.
The method builds on the possibility to control the nonlinear part of the mechanical potential landscape and introduce it periodically, adjusted in time with the mechanical harmonic oscillations.
Such periodic application inhibits the effect of the free motion and the restoring force terms in the Hamiltonian and allows approaching the state arising from the nonlinear potential only.
This is achieved similarly to how a stroboscopic measurement enables a quantum non-demolition detection of displacement~\cite{braginskyoptimal1978,cavesmeasurement1980}.
To prove feasibility of the method, we theoretically investigate realistic dynamics of a levitated nanoparticle in presence of simultaneously a harmonic and a strong, stroboscopically applied, nonlinear potentials enabled by the engineering of the trapping beam.
To run numerical simulations, we advance the theory of optomechanical systems beyond the covariance matrix formalism appropriate for Gaussian states.
Using direct Fock-basis and Suzuki-Trotter~\cite{suzukigeneralized1976} simulations we model the simultaneous action of the nonlinear potential and harmonic trap, and obtain the Wigner functions of the quantum motional states achievable in this system.
We predict very nonclassical negative Wigner functions~\cite{bartlettuniversal2002,ghosenongaussian2007} generated by highly nonlinear quantum-mechanical evolution for time shorter than one mechanical period.
The oscillations of Wigner function reaching negative values, in accordance with estimates based on unitary dynamics, witness that the overall quantum state undergoes unitary transformation $\exp(\ii V(\ox)\tau)$ sufficient for universal quantum processing~\cite{lloydquantum1999}.
To justify it, we prove a nonlinear combination of the canonical quadratures of the mechanical oscillator to be squeezed below the ground state variance which is an important prerequisite of this state being a resource for the measurement-based quantum computation~\cite{miyataimplementation2016,marekgeneral2018}.
For numerical simulations, we focus our attention to realistic version of the key nonlinearity, namely the cubic one with $V(x) \propto x^3$, and find good agreement of the predictions based on experimentally feasible dynamics with the lossless and noiseless unitary approximation.
The method allows straightforward extension to more complex nonlinear potentials which can be used for flexible generation of other resources for nonlinear gates and their applications~\cite{sefihow2011,marekgeneral2018}.
In comparison with simultaneously developed superconducting quantum circuits~\cite{sivakkerrfree2019}, an advantage of our approach stems from a much larger flexibility of nonlinear potentials.
Stroboscopic driving of an optomechanical cavity in a linear regime was considered in~\cite{brunellistroboscopic2020} for the purpose of cooling and Gaussian squeezing of the mechanical mode.

\section{Results} 
\label{sec:results}

\begin{figure}[h!]
	\centering
	\includegraphics[width=\linewidth]{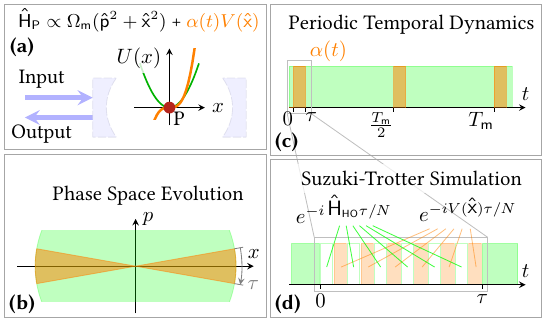}
	\caption{%
        Scheme of the proposed stroboscopic method.
		(a) A levitated optomechanical system as an illustration of mechanical oscillator in a nonlinear potential.
		A dielectric subwavelength particle (P) is trapped by a tweezer (not shown).
        The particle feels a total potential $U(x) = \Omega\s{m} x^2/ 4 + \alpha(t) V(x)$ that is a sum of the quadratic (green) and the nonlinear (orange, here: cubic) parts, both provided by the trapping beam.
		The particle can be placed inside a high-$Q$ cavity and probed by the laser light.
		(b,c) Stroboscopic application of nonlinear potential.
        The nonlinear part of the potential is switched on for only a short fraction of the mechanical period (orange segments).
        The quadratic trapping potential (green segments) is present throughout all the evolution.
		(d) Suzuki-Trotter simulation of stroboscopic evolution of the mechanical mode.
		In the figure, orange segments represent action of the nonlinear potential, empty and filled green segments correspond, respectively to unitary and damped harmonic evolution.
	}
	\label{fig:0}
\end{figure}

\subsection{The nonlinear stroboscopic protocol}
To implement the stroboscopic method, it is possible to versatilely use a levitated nanoparticle~\cite{konopik_nonequilibrium_2020} with optical~\cite{meyerresolvedsideband2019}, electric~\cite{goldwaterlevitated2019} or magnetic~\cite{vinanteultralow2020} trapping.
It is also possible to use a mirror equipped with a fully-optical spring~\cite{corbittalloptical2007}, or a membrane with electrodes allowing its nonlinear actuation and driving~\cite{sridaranelectrostatic2011}.
In any of such systems, the mechanical mode can be posed into nonlinear potential $V (x)$, particularly, the cubic potential $V_3 (x) \propto x^3$ for the pioneering test.
In this manuscript we focus on the experimental parameters peculiar to the levitated nanoparticles~\cite{deliccavity2019,deliccooling2020}, although the principal results remain valid for the other systems as well.
We also focus here solely on the evolution of the mechanical mode of the optomechanical cavity, assuming that the coupling to the optical cavity mode (blue in~\cref{fig:0}) is switched off.

The mechanical mode is a harmonic oscillator of eigenfrequency $\Omega\s{m}$, described by position and momentum quadratures, respectively, $\oper x$ and $\oper p$, normalized such that $\comm{ \oper x }{\oper p} = 2 \ii$.
The oscillator is coupled to a thermal bath at rate $\damp$.
We also assume fast stroboscopic application of an external nonlinear potential $\alpha(t) V(x)$ with a piecewise constant $\alpha(t)$ illustrating the possibility to periodically switch the nonlinear potential on and off as depicted at~\cref{fig:0}(a).
The Hamiltonian of the system, therefore, reads~($\hbar = 1$)
\begin{equation}
	\label{eq:hamiltonian}
	\oper H = \oper H\s{HO} + \alpha(t) V (\oper x),\quad  \oper H\s{HO} = \frac 1 4 \Omega\s{m} ( \oper x^2 + \oper p^2 ),
\end{equation}

To illustrate the key idea behind the stroboscopic method, we first examine the regime of absent mechanical damping and decoherence.
In this case, the unitary evolution of the oscillator is given by $\rho(t) = \un (t,t_0) \rho(t_0) \un^\da (t,t_0)$, with $\un (t + \delta t, t) = \exp[ - \ii \oper H \delta t ]$.
When the nonlinearity is switched on permanently ($\alpha(t) = \alpha_0$), the free evolution dictated by $H\s{HO}$ mixes the quadratures of the oscillator which prohibits the resulting state from possessing properties of the target nonlinear quantum state arising purely from $V(x)$ regardless of the nonlinearity strength (see Supplementary Note 4 
for more details).
Willing to obtain a unitary transformation as close to $\exp[ - \ii V( \oper x) \tau]$ as possible despite a constant presence of $\oper H\s{HO}$, we assume that the nonlinearity is repeatedly switched on for infinitesimally short intervals of duration $\tau$.
If the duration $\tau$ is sufficiently short for the harmonic evolution to be negligible, the resulting evolution during $\tau$ is approximately purely caused by the $V(x)$ part of the Hamiltonian.
To enhance the magnitude of the effect of the nonlinear potential, $V (\ox)$ we can apply it every $2 \pi / \Omega\s{m}$ for short enough intervals as shown in~\cref{fig:0}~(b,c).
This allows to establish an effective rotating frame within which the nonlinearity is protected from the effect of harmonic evolution.
Realistically, the stroboscopic application corresponds to $\alpha(t) = \sum_k \delta_\tau ( t - 2 \pi  k/ \Omega\s{m}),$ with $k \in \mathbb Z$, where $\delta_\tau$ is a physical approximation of Dirac delta function with width $\tau$ much shorter than the period of mechanical oscillations: $\tau \Omega\s{m} \ll 1$.
Then we can consider the evolution over a number $M$ of harmonic oscillations as consisting of subsequent either purely harmonic or purely nonlinear steps, and the evolution operator can be approximately written as
\begin{equation}
	\label{eq:instantaneous}
	\un (t + M T\s{m}, t) = \left[ \un (t + T\s{m}, t ) \right]^M
	\approx \left\{ \exp[ - \ii \oper H\s{HO} T\s{m} ] \exp[ - \ii V ( \oper x ) \tau ] \right\}^M = \exp[ - \ii M V (\oper x ) \tau ].
\end{equation}
For the last equality we used the fact that the unitary harmonic evolution through a single period of oscillations is an identity map: $ \un\s{HO} (t + T\s{m}, t ) \equiv \exp[ - \ii \oper H\s{HO} T\s{m}]~=~\oI$.
Motion of a real mechanical oscillator can be approximated by a harmonic unitary evolution with good precision because of very high quality of mechanical modes of optomechanical systems~\cite{masoncontinuous2019,windeycavitybased2019,deliccavity2019}.
\cref{eq:instantaneous} shows that the effect of sufficiently short pulses of the strong nonlinear potential timed to be turned on precisely once per a period of mechanical oscillations $M$ times, is equivalent to an $M$-fold increase of the nonlinearity.

A further improvement is possible by noting that undamped harmonic evolution over a half period simply flips the sign of the two quadratures $(\oper x, \oper p) \mapsto - (\oper x , \oper p)$.
Therefore, it is possible to similarly apply the potential twice per period, switching its sign each second time.
This can be formalized as setting $\alpha (t) = \sum_k ( -1 )^k \delta_\tau ( t - \pi k / \Omega\s{m} ), \ k~\in~\mathbb Z$.
In this case,
\begin{equation}
	\un (t + T\s{m} , t )
	=
	\ee^{  - \ii \oper H\s{HO} T\s{m}/2  } \ee^{  + \ii V (\oper x ) \tau  }
	\ee^{  - \ii \oper H\s{HO} T\s{m} /2   }
	\ee^{  - \ii V (\oper x ) \tau  } = \ee^{  - 2 \ii V (\oper x ) \tau  },
\end{equation}
and therefore, after $M$ periods
\begin{equation}
	\label{eq:mperiods}
	\un ( t + M T\s{m} , t )  = \exp[ - \ii \cdot 2 M \cdot V ( \oper x ) \tau ].
\end{equation}

The idealised scheme proposed above in reality faces two potentially deteriorative factors: the finiteness of the duration of the nonlinearity $\tau$, and the mechanical decoherence caused by the thermal environment.
We take a proper account of these two factors by considering the evolution as consisting of two kinds: (i) unitary undamped dynamics in a sum of the quadratic and nonlinear potentials, (ii) damped harmonic evolution between those.
These two kinds of evolution are subsequently repeated, as shown in~\cref{fig:0}~(b,c).
We develop an advanced method based on Suzuki-Trotter simulation (STS) to simulate the quantum state of a realistic optomechanical system after the application of our proposed protocol.
It is worth noting that the STS is typically used to approximately achieve a novel evolution $\un$ from experimentally available exact unitaries~\cite{lamatadigitalanalog2018,lougovski_digitalanalog_2020}.
In our work, we use STS to simulate and justify that the experimentally available compound evolution $\un$ can approximate one of its building blocks~$\un\s{NL} ( \delta t ) \equiv \exp( - \ii V (x) \delta t )$.
To verify the convergence of STS we also perform simulations in the Fock-state basis that allow direct computation of the propagator corresponding to the Hamiltonian~\eqref{eq:hamiltonian}.
Fock-state-basis simulations unfortunately do not grant access to phase-space distributions such as Wigner function~\cite{weinbubrecent2018} which makes use of STS the primary strategy.
Excellent agreement between the very distinct methods (STS and Fock-state-basis simulations) indicates that our results are correct.
The details of the simulation methods are presented in Supplementary Notes 1 and 2.

We omit damping and thermal decoherence during the fast unitary action of the combined potential.
Such applications are assumed to happen every half of a mechanical oscillation and have durations much shorter than the mechanical period, therefore, due to high quality of state-of-the-art mechanical oscillators, such omission is justified.
The joint action is simulated using STS and verified by Fock-state-basis simulation.
Between the applications of the nonlinear potential $V(x)$ the mechanical oscillator experiences damped harmonic evolution described by the linear Heisenberg-Langevin equations
\begin{equation}
	\dot x = \Omega\s{m} p;
	\quad
	\dot p = - \Omega\s{m} x - \damp p + \sqrt{ 2 \damp } \xi,
\end{equation}
where $\xi$ is the quantum Langevin force, obeying $\comm{ \xi(t) }{ x (t) } = \ii \sqrt{ 2 \damp }$ and $ \frac 1 2 \avg{\left\{ \xi(t) , \xi(t') \right\}} = ( 2 n\s{th} + 1 ) \delta ( t - t')$ with $n\s{th}$ being the mean occupation of the bath.
The experimentally accessible value of the heating rate $H\s{m}$ is given by $H\s{m} = \damp n\s{th}$.
The density matrix $\rho (T\s{m}/2)$ of the particle after half of a period of oscillations including the action of the nonlinear potential and subsequent damping can be formally written as
\begin{equation}
	\label{eq:singleperiod}
\rho(\tau) = \mathcal N \left[ \rho (0) \right] = \mathcal D \left[ \un (\tau , 0 ) \rho (0) \un^\da ( \tau , 0 ) \right].
\end{equation}
where
$\un(\tau,0)$ describes the particle's unitary dynamics in combined potential, 
and $\mathcal D [\ \bullet\ ]$ indicates the mapping performed by the damping.
Generalization of~\eqref{eq:singleperiod} to include the second half of the period, and the subsequent generalization to multiple periods, is straightforward.

Using these advanced numerical tools, further elaborated in~\cref{sec:methodsofnumericalcomputations} we evaluate the quantum state of the mechanical oscillator after the protocol and explore the limits of the achievable nonlinearities in the optomechanical systems that are accessible now or are within reach.

\subsection{Application to the cubic nonlinearity}

\begin{figure}[htb]
	\centering
	\includegraphics[width=.64\linewidth]{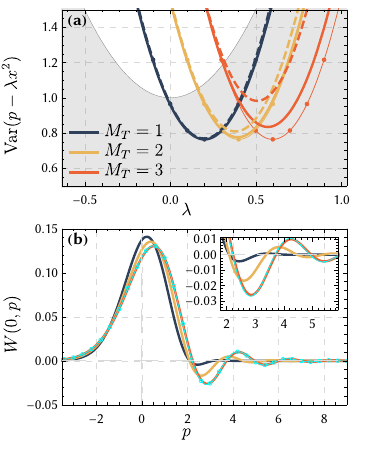}
	\caption{Results of the simulations. Results of stroboscopic application of cubic nonlinear potential $V= \gamma x^3 / ( 6  \tau )$ to the squeezed thermal state (initial occupation $n_0 = 0.05$, squeezing $s = 1.2$) over multiple ($M\s{T } = 1,2$ or $3$) halves of mechanical periods.
	(a) Squeezing of nonlinear quadrature.
	Thick lines correspond to stroboscopic method with realistic parameters and thermal noise, thin lines, to purely unitary application of the nonlinearity.
    Dashed lines correspond to the result of evolution driven by the full Hamiltonian $\oper H$ \eqref{eq:hamiltonian} from the same initial state over time $\tau, 2\tau$ and $3\tau$ respectively.
	(b) Wigner function of motional states.
	Solid lines correspond to the same states from (a).
	Cyan line with markers (overlapping with solid red line) shows the unitary application of nonlinearity.
	Other parameters are: potential stiffness $\gamma = \gamma_0 = 0.2$, application duration $\tau \Omega\s{m} = \Theta = \Theta_0 = \pi/100$, environmental thermal heating rate parameter $H_{0} = 4 \pi H\s{m}/ \Omega\s{m} =  0.002$.
}%
	\label{fig:1:application}
\end{figure}

Motivated by the role of cubic nonlinearity for the universal continuous-variable quantum processing, we illustrate the devised method by numerically evaluating the evolution of a levitated particle under a stroboscopic application of a cubic potential $V(x) \propto x^3$.
The nonlinear phase gate $\ee^{ - \ii V(x)}$ is a limit case of motion, it modifies only momentum of the object without any change of its position.
This nondemolition aspect is crucial for use in universal quantum processing.
A nonlinear phase state (particularly, the cubic phase state introduced in~\cite{gottesmanencoding2001}) as the outcome of evolution of the momentum eigenstate $\ket{ p = 0}$ in a nonlinear potential $V(x)$ is defined as
\begin{equation}
	\label{eq:GKP}
	\ket{ \nlgain _V} \propto \int \dd x \ee^{ \ii V(x ) \tau } \ket{ x },
\end{equation}
where $V(x )$, is a highly nonlinear potential, and $\ket{ x }$ the position eigenstate $\oper x \ket x = x$.
The state~\eqref{eq:GKP} requires an infinite squeezing possessed by the ideal momentum eigenstate before the nonlinear potential is applied.
More physical is an approximation of this state obtained from a finitely squeezed thermal state $\rho_0 (r, n_0)$, ideally, vacuum,
by the application of $V$:
\begin{equation}
	\label{eq:cubic}
	\rho ( V , r, n_0 )  = \ee^{ \ii V(\oper x) \tau } \rho_0 (r , n_0) \ee^{ - \ii V( \oper x) \tau }.
\end{equation}
The initial state $\rho_0$ is the result of squeezing a thermal state with initial occupation $n_0$
\begin{equation}
	\rho_0 (r , n_0) = \oper S(r) \rho_0 ( 0 , n_0) \oper S^\da (r)
\end{equation}
where $ \oper S(r) = \exp[ \tfrac 12 r^* (\oper a)^2 - \tfrac 12 r ( \oper a^\da )^2] $ is a squeezing operator ($\oper a = (\oper x + \ii \oper p)/2$), and the initial state $\rho_0$ is thermal with mean occupation~$n_0$.
Phase of the squeezing parameter $r = |r| \ee^{ \ii \theta }$ determines the squeezing direction.
When $n_0 = 0$, $|r| \to +\infty$, and $\theta = \pi$, the initial state is infinitely squeezed in $p$, and ~\cref{eq:cubic} approaches the ideal cubic state~\eqref{eq:GKP}.

The quantum state obtained as a result of the considered sequence of interactions approximates the ideal state given by~\cref{eq:GKP}.
The quality of the approximation can be assessed by evaluating the variance of a nonlinear quadrature $p - \vk x^2$, or the cuts of the Wigner functions of the states.
A reduction in nonlinear quadrature variance below the vacuum is a necessary condition for application of these states in nonlinear circuits~\cite{miyataimplementation2016,marekgeneral2018}.
On the other hand, the phase-space interference fringes of the Wigner function reaching negative values are a very sensitive witness of quantum non-Gaussianity of the states used in the recent experiments~\cite{yoshikawacreation2013,kienzlerobservation2016,wangschrodinger2016,johnsonultrafast2017}.
Fidelity happens to be an improper measure of the success of the preparation of the quantum resource state~\cite{yukawaemulating2013} because it does not predict either applicability of these states as resources or their highly nonclassical aspects.


A noise reduction in the cubic phase gate can be a relevant first experimental test of the quality of our method.
The approximate cubic state obtained from a squeezed thermal state (that is, the state~\eqref{eq:cubic} with $V(x) = \gamma x^3 /( 6 \tau )$) should possess arbitrary high squeezing in the variable $p - \vk x^2$ for $n_0 = 0$ given sufficient squeezing of the initial mechanical state.
The state~\eqref{eq:cubic} obtained from a squeezed in momentum state, has the following variance of the nonlinear quadrature
\begin{equation}
	\label{eq:nlvariance}
	\sigma_3 ^\rho (\vk) \equiv \Tr(\rho  ( \op - \vk \ox^2 )^2 ) - \left( \Tr \rho ( \op - \vk \ox^2 ) \right)^2
	= \frac{ v\s{th} }{ s^2 } + 2 ( \vk - \gamma )^2 ( s^2 v\s{th} )^2,
\end{equation}
where $v\s{th} = 2 n_0 + 1 $ is the variance of each canonical quadrature in the initial thermal state before squeezing, and $s = \ee^{r }$ is the magnitude of squeezing.
An important threshold is the variance of the nonlinear quadrature attained at the vacuum state ($n_0 = 0$, $s = 1$)
\begin{equation}
	\label{eq:nlvarvacuum}
	\sigma_3\up{vac} (\vk) = 1 + 2 \lambda^2 .
\end{equation}
Application of a unitary cubic evolution to the initial vacuum state displaces this curve along the $\vk$ axis by $\gamma$.
Further, as follows from~\cref{eq:nlvariance}, squeezing the initial state allows reducing the minimal value of $\sigma_3$, and increase of the initial occupation $n_0$ causes also increase of $\sigma_3$.
Suppression of fluctuations in the nonlinear quadrature is a convenient figure of merit because it is a direct witness of the applicability of the quantum state as a resource for measurement-based quantum information processing~\cite{miyataimplementation2016,marekgeneral2018} and a witness of non-classicallity~\cite{mooreestimation2019}.
It can be evaluated in an optomechanical systems with feasible parameters using pulsed optomechanical interaction~\cite{mooreestimation2019} without a full quantum state tomography.

In~\cref{fig:1:application}~(a) we show the variance $\sigma_3$ at the instants $t = M\s T T \s m/2$.
Each of the curves takes into account also the interaction with the thermal environment lasting $T \s m/2$ after each application of the nonlinearity.
The heating rate parameter $H_0 = 4\pi H\s{m}/\Omega\s{m}$ assumes value $H_0 = 2 \times 10^{-3}$.
For an oscillator of eigenfrequency $\Omega\s{m} = 2 \pi \times \SI{100}{\kHz}$ and $Q=10^6$ is equivalent to occupation of the environment equal to $n\s{th} \approx 10^{7}$ phonons.
This is the equilibrium occupation of such an oscillator at the temperature of \SI{50}{\K}.
A recent experiment of ground state cooling of a levitated nanoparticle~\cite{deliccooling2020} reported the heating rate corresponding to $H \approx 10^2 H_0$.
A proof of robustness of the method against such heating is in the Supplementary Note 3.

Thin lines with markers show the analytic curves defined by~\cref{eq:nlvariance} for the corresponding values of $\gamma$.
The good quantitative correspondence between the approximate states resulting from the realistic stroboscopic application of nonlinear potential and the analytic curves again proves the validity of the stroboscopic method.
Importantly, each of the curves has an area where it lies below the corresponding ground-state level~$\sigma_3\up{vac}$.
This means each of the corresponding states gives advantage over vacuum if used as ancilla for the cubic gate.
The dashed lines show the simulated quantum states obtained from the same initial state by longer unitary evolution according to the full Hamiltonian from~\cref{eq:hamiltonian}, that is $\ee^{ - \ii \oper H n \tau } \rho \ee^{ \ii \oper H n \tau }$ where $n = 1,2,3$ for blue, yellow and red correspondingly.
Further divergence from ideal than that of the ones corresponding to the stroboscopic method witnesses an advantage of the latter in generation of the resource for the measurement-based computation.

The stroboscopic application of a fixed limited gain nonlinearity, therefore, indeed allows amplification of nonlinearity in accordance with~\cref{eq:mperiods}.
Importantly, even despite requiring longer evolution in a noisy environment, the stroboscopic method allows better amplification than a unitary longer application of the nonlinearity in presence of the free evolution ($\propto \oper p^2$) and harmonic ($\propto \oper x^2$) terms in the Hamiltonian.

The non-Gaussian character of the prepared quantum state can be witnessed via its Wigner function $W (x,p)$ which for a quantum state $\rho$ reads~\cite{schleichquantum2011}
\begin{equation}
	W (x,p) = \frac{ 1 }{ 2 \pi } \int \limits_{- \infty}^{\infty} \dd y \ee^{ - \ii p y } \mel{ x + y }{ \rho }{ x - y }.
\end{equation}
Wigner function (WF) shows a quasiprobability distribution over the phase space spanned by position $x$ and momentum $p$ and its negativity is a prerequisite of the non-classicallity of a quantum state.
In~\cref{fig:1:application}~(b) are Wigner functions $W (0,p)$ of the mechanical oscillator computed for the same approximate states as in the panel~(a).
The Wigner function of an ideal cubic phase state, i.e. the state given by~\cref{eq:GKP} for $V(x) = x^3 \nlgain/ (6 \tau) $, reads~\cite{gottesmanencoding2001,ghosenongaussian2007}
\begin{equation}
	W\s{CPS} (x,p) \propto \Ai \left[ \left(\frac{ 4 }{ \nlgain } \right)^{1/3} \left( \nlgain x^2 - p \right) \right],
\end{equation}
where $\Ai(x)$ is the Airy function.
This function with apparent non-Gaussian shape in the phase space exhibits fast oscillations reaching negative values in the positive momentum for any $\nlgain >  0$.
The Wigner functions of the states obtained by application of the stroboscopic protocol approach the one of~\cref{eq:cubic}.
Each of them exhibits areas with negative values which proves quantum non-Gaussian character of the resulting state.
Moreover, with increased number of stroboscopic applications involved, the resulting approximate state corresponds to stronger nonlinearity.
For the last curve where $M\s T = 3$ we also show the result of an idealized instantaneous unitary application of an equal total nonlinearity by a cyan line with markers which is indistinguishable from the line for the approximate evolution.
For this pair of curves we also have an estimate for the overlap $\Tr[ \rho\s{red} \rho\s{cyan} ] / \Tr[ \rho\s{cyan}^2 ] = 0.9877$.
Despite an excellent overlap of the ideal and approximate Wigner functions, the important resource, squeezing $\sigma_3$ shows a noticeable deviation from the ideal scenario.
It is for this reason that we choose the squeezing $\sigma_3$ as the main figure of merit for the protocol.


\section{Discussion}

In this article we have proposed and theoretically analyzed a protocol to create a nonlinear motional state of the mechanical mode of an optomechanical system with controllable nonlinear mechanical potential.
The method uses the possibility to apply the nonlinear potential to the motion of mechanical object in a stroboscopic way, twice per a period of oscillations.
This way of application allows reducing the deteriorative effect of the free oscillations and approach the effect of pure action of nonlinear potential.
In contrast to other methods of creating non-classical states by a continuous evolution in presence of nonlinear terms in the Hamiltonian~\cite{ballentine_moment_1998,brizuela_classical_2014,brizuela_statistical_2014}, our method allows approaching the states that approximate evolution according to the unitary~$\ee^{ \ii V(x) }$ where $V(x)$ is the nonlinear potential profile.
We tested our method on a cubic nonlinearity $\propto \ox^3$ though the method is applicable to a wide variety of nonlinearities.
Our simulations prove that application of the protocol allows one to obtain the squeezing in the nonlinear quadrature below the shot-noise level even if the initial state of the particle is not pure.
The nonlinear state created in a stroboscopic protocol clearly outperforms as a resource the vacuum for which the bound~\eqref{eq:nlvarvacuum} holds.
Moreover, the stroboscopic states approximate the one defined by~\cref{eq:cubic} obtained in absence of the free rotation and thermal decoherence.
We also verify that the stroboscopic application of the cubic nonlinear potential generates nonclassical states under conditions that are further from optimal than the ones of~\cref{fig:1:application}.
In particular, we find that the heating rate $H\s{m}$ can be increased approximately 100-fold until the curve $\sigma_3 (\lambda)$ fails to overcome the vacuum boundary $1 + 2 \lambda^2$.
We are able to prove that when the duration of the application $\tau$ is increased tenfold such that the corresponding product $V(x) \tau$ remains constant, (that is, a less stiff potential is applied stroboscopically for proportionally longer time intervals), the resulting nonlinear state as well shows squeezing in $\sigma_3$.
This proves robustness of the method to the two major imperfections.

We have shown the method to work for the parameters inspired by recent results demonstrated by the levitated optomechanical systems~\cite{kieselcavity2013,magrininearfield2018}.
The optical trap with a cubic potential has been already used in the experiments~\cite{silerthermally2017,silerdiffusing2018}.
Levitated systems~\cite{millen_quantum_2020}, including electromechanical systems~\cite{martinetzquantum2020}, have recently shown significant progress in the motional state cooling~\cite{windeycavitybased2019,deliccavity2019,deliccooling2020} and feedback-enhanced operation~\cite{vovroshparametric2017} which lays solid groundwork of the success of the proposed protocol.
The authors of Ref.~\cite{silerdiffusing2018} report the experimental realisation of the potential $V(X) = \mu X^3/6$ with $\mu \approx 8 k\s{B} T \SI{}{\micro\meter^{-3}}$, where $X = x \sqrt{ \hbar / ( 2 m \Omega\s{m} )}$ is the dimensional displacement of the oscillator, $m$ is its mass, $k\s{B}$ is the Boltzmann constant and $T$ temperature.
From this value we can make a very approximate estimate for the nonlinear gain $\gamma = \mu ( \hbar / ( 2 m \Omega\s{m} ) )^{3/2} \tau \approx 1.2 \times 10^{-3}$ assuming duration $\tau = \pi/ (50 \Omega\s{m})$, temperature  $T = \SI{300}{\kelvin}$, frequency $\Omega\s{m} = 2 \pi \times \SI{1}{\kilo\hertz}$, and a mass $m = \SI{4e-15}{\gram}$ of a silica nanoparticle of $\SI{70}{\nano\meter}$ radius.

Experimental implementation of the proposed method can guarantee preparation of a strongly nongaussian quantum motional state.
Further analysis of such a state will require either a full state tomography or better suited well-tailored methods to prove the nonclassicality~\cite{vanneroptomechanical2015}.
An analysis of the estimation of the non-linear mechanical quadrature variance via pulsed optomechanical interaction can be found in~\cite{mooreestimation2019}.
The optical read out can be improved using squeezed states of light~\cite{filiptransfer2015}.
This experimental step will open applications of the proposed method to other nonlinear potentials relevant for quantum computation~\cite{sefihow2011,marekdeterministic2011,miyataimplementation2016,marekgeneral2018}, quantum thermodynamics~\cite{zhangquantum2014,dechantalloptical2015} and quantum force sensing~\cite{rugarsingle2004, degennanoscale2009}.

In our simulations we focused solely on the dynamics of the mechanical mode and assumed the conventional optomechanical interaction absent.
This interaction, well developed in recent years, provides a sufficiently rich toolbox that allows incorporation of the mechanical mode into the optical circuits of choice~\cite{aspelmeyercavity2014}.
As an option, a prepared nonlinear state can be transferred to traveling light pulse~\cite{filiptransfer2015} using optomechanical driving.
In a more complicated scenario, one can add optomechanical interaction to the stroboscopic evolution to obtain even richer dynamics.
A complete investigation of such dynamics, however, goes beyond the scope of the present research focused on the preparation of nonlinear motional states.

In parallel with the experimental verification, the stroboscopic method can be used to analyse other higher-order mechanical nonlinearities such as $V(x)\propto x^4$ or tilted double-well potentials required for tests of recently disputed quantum Landauer principle~\cite{millerquantum2020}, counter-intuitive Fock state preparation~\cite{simontrappedion2020} and approaching macroscopic superpositions~\cite{abdidissipative2016}.
\section{Methods} 
\label{sec:methodsofnumericalcomputations}

\subsection{Tools of numerical simulations}

The~\cref{eq:singleperiod} approximates the damped evolution of an oscillator in a nonlinear potential by a sequence of individual stages of harmonic, nonlinear and damped harmonic evolution (see~\cref{fig:0}).
Below we elaborate on how it is possible to simulate such dynamics using Wigner function in the phase space, density matrix in position, momentum and Fock basis.

First, we evaluate the action of the nonlinearity during the stroboscopic pulse.
While the nonlinearity is on, the Hamiltonian of the system reads
\begin{equation}
	\label{eq:hamsplit}
	\oper H = \oper H\s{HO} + V(\oper x) = \oper H_p + \oper H_x,
\end{equation}
where
$\oper H_p = \frac{ \Omega\s{m} }{4 } \oper p^2, $  and
$\oper H_x = \frac{ \Omega\s{m} }{4 } \oper x^2 + \frac{ \gamma }{ 6 } \oper x^3.$

Our important simulation tool is the Suzuki-Trotter simulation (STS, see Ref.~\cite{suzukigeneralized1976}) for $\un$
\begin{equation}
	\un (t + \tau, t ) = \Big[ \exp[ - \ii ( \oper H\s{HO} + V( \ox ) ) \tfrac \tau N ] \Big]^N
	= \Big[ \un\s{HO} (\tfrac \tau N) \un\s{NL} (\tfrac \tau N ) + O \left( \left( \tfrac{ \tau }{ N } \right)^2 \right) \Big]^N ,
	\label{eq:suzuki-trotter}
\end{equation}
where $\un\s{HO} (\delta t) \equiv \exp( - \ii H\s{HO} \delta t )$, $\un\s{NL} ( \delta t ) \equiv \exp( - \ii V (x) \delta t )$,
$N$ is called the Trotter number.
The accuracy of the approximation is, thereby, increasing with decreasing $\tau /N$.
Despite $\tau$ being now sufficiently large to take into account noticeable free rotation through an angle $\Omega\s{m} \tau$ in the phase space, we still assume that $\tau$ is much shorter than the mechanical decoherence timescale, set by the heating rate $H\s{m}$.
This is well justified for the current experiments~\cite{jaindirect2016,deliccavity2019,deliccooling2020}, also see Supplementary Note 2. 
The STS requires the summands forming the Hamiltonian to be self-adjoint which is not always the case of $V(x)$, in particular $V(x) \propto x^3$.
We take the necessary precautions by considering such nonlinearities over only short time in a finite region of the phase space.
Thus we cautiously take care of the quantum motional state being limited to this finite region.
To further verify the correctness of numerics via STS, we cross-check it using numerical simulations in Fock-state basis.

\cref{eq:hamsplit} shows the two possibilities to split the full Hamiltonian into summands to use the STS.
We use these two possibilities to compute independently the mechanical state in order to verify the correctness of the STS in 
Supplementary Note 1.

First, we start from a squeezed thermal state $\rho(0)$, which has a representation by the Wigner function in the phase space
\begin{equation}
	\label{eq:wigsqtherm}
	W\s{th} ( x, p  ; n_0, s ) =
	\frac{ \exp \left( - \frac 1 2 \left[ \frac{ ( x / s )^2 + ( p s )^2 }{ 2 n_0 + 1 } \right] \right)}{ 2 \pi ( 2 n_0 + 1 ) }.
\end{equation}
The Wigner function corresponding to a quantum state $\rho$ is defined~\cite{schleichquantum2011} as
\begin{equation}
	W (x,p) = \frac{ 1 }{ 2 \pi } \int \limits_{- \infty}^{\infty} \dd y \ee^{ - \ii p y } \mel{ x + y }{ \rho }{ x - y },
\end{equation}
and the corresponding density matrix element can be obtained from the Wigner function by an inverse Fourier transform.
It is therefore possible to extend this approach to any $W(x,p)$ beyond the Gaussian states.

The evolution of a state $\rho$ under action of a Hamiltonian proportional to a quadrature $\oper q$ can be straightforwardly computed in the basis of this quadrature, where it amounts to multiplication of density matrix elements with $c$-numbers:
\begin{equation}
	\label{eq:unitary-in-quad-basis}
	\bra{ q } \ee^{ - \ii \oper H_q (\oper q)  t } \rho \ee^{ \ii \oper H_q (\oper q) t } \ket {q'} =
	\mel{q }{\rho}{q'} \ee^{ - \ii ( H_q (q) - H_{q'} (q') ) t }.
\end{equation}
In particular, the nonlinear evolution reads
\begin{equation}
	\bra{ x } \un\s{NL} ( \delta t ) \rho \un^\da\s{NL} (\delta t) \ket{ x'} = \bra x \rho \ket {x'} \ee^{  - \ii [ V(x) - V(x')] \delta t  }.
\end{equation}

The undamped harmonic evolution driven by $\oper H\s{HO}$ can be represented by the rotation of Wigner function (WF) in the phase space.
A unitary rotation through an angle $\theta = \Omega\s{m} \delta t$ in the phase space maps the initial WF $W(x,p; t)$ onto the final $W(x, p; t + \delta t)$ as
\begin{equation}
	\label{eq:wigner-rotation}
	W (x,p; t + \delta t) = W( x \cos \theta - p \sin \theta , p \cos \theta + x \sin \theta; t ).
\end{equation}

The unitary transformation of the density matrix can be as well computed in the Fock state basis.

Finally, damped harmonic evolution of a high-Q harmonic oscillator over one half of an oscillation can also be evaluated in the phase space as a convolution of the initial Wigner function $W(x,p;t)$ with a thermal kernel
\begin{equation}
	\label{eq:thermalwigner}
    W (x , p ; t + \frac{ \pi }{ \Omega } ) = \iint \limits_{-\infty }^{ \infty } \dd u \dd v W_i (x - u , p - v ; t ) W\s{B} (u , v ),
\end{equation}
where the expression for the kernel reads
\begin{equation}
	W\s{B} (u,v) = \frac{ 1 }{ 2 \pi \sigma\s{th}} \exp \left[ - \frac{ u^2 + v^2 }{ 2 \sigma\s{th}} \right],
\end{equation}
with $\sigma\s{th} = ( 2 n\s{th} + 1 ) { 2 \pi \damp}/{ \Omega\s{m} }$, where $n\s{th} \approx k\s{B} T / ( \hbar \Omega\s{m} )$ is the thermal occupation of the bath set by its temperature $T$.
In terms of the heating rate $H\s{m}$, $\sigma\s{th} = 4 \pi H\s{m} / \Omega\s{m}$.
\cref{eq:thermalwigner} is obtained by solving the joint dynamics of the oscillator and bath followed by tracing out the latter.
The detailed derivation of~\cref{eq:thermalwigner} can be found in 
Supplementary Note 2.

Using these techniques, one can evaluate the action of the map $\mathcal N$ defined by~\cref{eq:singleperiod} on the state of the quantum oscillator.
This yields the quantum state of the particle after one half of a mechanical oscillation.
Repeatedly applying the same operations, one can obtain the state after multiple periods of the mechanical oscillations.
Our purpose is then to explore the limits of the achievable nonlinearities in optomechanical systems that are accessible now or are within reach.


\section*{Data Availability}
\label{sec:dataavailability}

The datasets generated and analyzed during the current study are available from the
corresponding author on reasonable request.

\begin{DIFnomarkup}
\end{DIFnomarkup}

\section*{Acknowledgments}
	A.A.R. acknowledges the support of the project 20-16577S of the Czech Science Foundation and national funding from the MEYS under grant agreement No. 731473 and from the QUANTERA ERA-NET cofund in quantum technologies implemented within the European Union's Horizon 2020 Programme (project TheBlinQC).
    R.F. acknowledges the project 21-13265X of the Czech Science Foundation.
\section*{Author Contributions} 
\label{sec:authorcontributions}

R.F. conceived the theoretical idea.  A.A.R. performed the numerical computations.  Both authors participated in writing the manuscript.

\section*{Competing Interests} 
\label{sec:competinginterests}

The authors declare no competing interests.

\nocite{gerry_introductory_2004,palomakientangling2013,hallquantum2013}
\printbibliography

\clearpage

\SupplementaryMaterials

\begin{center}
	\textbf{\sffamily\LARGE Supplementary Information: \thetitle}
\end{center}


\section{Convergence of the Suzuki-Trotter approximation} 
\label{sec:convergenceofthesuzukitrotterapproximation}

In this section we numerically demonstrate that the Suzuki-Trotter approximation that we use to simulate the combined dynamics, indeed converges.
For this purpose we provide the plots of the variance of nonlinear variable $\Var( p - \lambda x^2 )$ computed in the approximate nonlinear state
\begin{equation}
	\rho (\tau) = \oper U (\tau, 0) \rho (0) \oper U^\da (\tau,0),
\end{equation}
where $\oper U (\tau , 0 )  = \exp[ - \ii \oper H \tau ]$ using different methods.

For the first two, we use the Suzuki-Trotter expansion of the unitary operator $\oper U$
\begin{equation}
	\exp[ - \ii ( \oper H_1 + \oper H_2 ) N \delta \tau ] \approx  \left(  \exp[ - \ii \oper H_1 \delta \tau ] \times \exp[ - \ii \oper H_2 \delta \tau ] \right)^N,
\end{equation}
where, depending on the method, we choose either
\begin{equation}
	\text{Method 1:} \quad \oper H_1 = \oper H_p, \quad \oper H_2 =  \oper H_x,
\end{equation}
or
\begin{equation}
	\text{Method 2:} \quad \oper H_1 = \oper H\s{HO}, \quad \oper H_2 = V (\ox),
\end{equation}
with notations of~\cref{eq:hamsplit}.
The action of individual unitary operators in Method 1 is computed in the corresponding quadrature basis according to~\cref{eq:unitary-in-quad-basis}.
In Method 2 the action of $\oper H_1$ is simulated by rotation of the Wigner function in the phase space according to~\cref{eq:wigner-rotation}, and the action of $\oper H_2$ in the position basis as in~\cref{eq:unitary-in-quad-basis}.
In the main text, Method 1 is used to produce the results.

For the Method 3, we write the Hamiltonian $\oper H$ in the Fock state basis.
This allows to directly compute matrix elements of $\oper U (\tau,0)$ in the Fock basis as well, without the need to use STS, and directly obtain the approximate nonlinear state.
This method, unfortunately, does not allow to obtain Wigner function directly.

Comparison of the nonlinear variances from different methods is in Supplementary~\cref{fig:convergence}.
We show that as the Trotter number $N$ increases, the simulation using Method 1 converges.
The similarly converging simulation using Method 2 is not shown to avoid cluttering of the figure.
Both methods approximately converge to the simulations in the Fock basis.
Method 2 shows larger deviation since the operation of rotation of the Wigner function in the phase space involves interpolation and is therefore less accurate.
This is the reason to choose Method 1 for the main text.
Good numerical coincidence of the results of all the three different methods proves convergence of the STS.

As an experiment, we also simulate the regime of operation that is accessible only to the levitated nanoparticles.
In such systems, one has in principle full control over the potential that the particle is exposed to, and can, therefore alternate between the quadratic trapping potential and the higher-order nonlinear one.
Formally this corresponds to having a Hamiltonian that, instead of~\cref{eq:hamiltonian}, reads
\begin{equation}
	\oper H = \frac{ \Omega\s{m} }{ 4 }\Big[ \op^2 +  ( 1 - \alpha (t)) \ox^2 \Big] + \alpha (t) V ( \ox).
\end{equation}
We also simulate the action of the nonlinear stage of this Hamiltonian, that is the action of $\oper H = \Omega\s{m} \op^2 /4 + \gamma \ox^3 / 6$, using Methods 1 and 3.
That is we simulate this regime using STS and Fock-state expansion.
The results of simulation show good agreement what again proves the STS convergence.
The nonlinear quadrature is squeezed stronger in this regime, which suggests that the strategy to switch between the potentials can be advantageous for the levitated nanoparticles, or other systems where a full control over the potential is available.

\section{Impact of the thermal noise}
\label{sec:impactofthethermalnoise}
Owing to recent progress in design and manufacturing of the nanomechanical devices, the thermal noise is strongly suppressed.
In this section we check the robustness of our scheme to the thermal noise from the environment and show that for the state-of-the art systems it is not hampering the quantum performance.

The thermal noise can be included by writing the standard Langevin equations:
\begin{equation}
	\dot x = \Omega\s{m} p;
	\quad
	\dot p = - \Omega\s{m} x - \damp p + \sqrt{ 2 \damp } \xi,
\end{equation}
where $\xi$ is the quantum Langevin force, obeying the commutation relation $\comm{ \xi(t) }{ x (t) } = \ii \sqrt{ 2 \damp }$ and markovian autocorrelation $ \frac 1 2 \avg{\left\{ \xi(t) , \xi(t') \right\}} = 2 n\s{th} + 1$.

The solution of these equations corresponding to one half of a period of the mechanical oscillations ($\tau\s{m} = \pi / \Omega\s{m}$) for the experimentally relevant regime of high-Q mechanical oscillator ($Q \equiv \Omega\s{m}/\damp \gg 1$) reads
\begin{align}
    x' & = \ee^{ - \damp \tau\s{m} / 2 } \Big[ ( \cos   \zeta  + \epsilon \sin \zeta ) x (0)
	                                           + \frac 1 \sigma \sin \zeta p (0) + \delta x (\tau\s{m})\Big],
	\label{eq:io:mech:term:x}
	\\
    p' & =  \ee^{ - \damp \tau\s{m} / 2 } \Big[ ( \cos  \zeta  - \epsilon \sin \zeta ) p (0)
	                                           + \frac 1 \sigma \sin \zeta x (0) + \delta p (\tau\s{m})\Big],
	\label{eq:io:mech:term:p}
\end{align}
with $\zeta = \Omega\s{m} \tau\s{m} \sigma$,  $\sigma = \sqrt{ 1 - ( \damp / 2 \Omega\s{m} )}$ and $\epsilon = \damp / ( 2 \Omega\s{m} \sigma) $.
The definitions for the noise operators $\delta x , \delta p$ read
\begin{align}
	\delta x (\tau\s{m}) & = \sqrt{ 2 \damp } \int_0^{\tau\s{m} } \dd t \ee^{ \damp t / 2 } \sin \Omega\s{m} \sigma ( \tau\s{m} - t ) \xi ( t),
	\\
	\delta p (\tau\s{m}) & = \sqrt{ 2 \damp } \int_0^{\tau\s{m} } \dd t \ee^{ \damp t / 2 } \Big[ \cos \Omega\s{m} \sigma ( \tau\s{m} - t )
						  - \epsilon \sin \Omega\s{m} \sigma (\tau\s{m} - t) \Big] \xi (t).
\end{align}
These are the canonical quadratures of a mode in a thermal state with variance $ \sigma\s{th} = 2 n\s{th} + 1$.

The transformation of the Wigner function can be found as follows.
Consider that at instant $t = 0$ the mechanical oscillator has WF $W\s{m}( x(0), p(0))$.
The mode of the bath is in a thermal state with WF
\begin{equation}
	W\s{B} (\delta x , \delta p ; \sigma\s{th}) = \frac{ 1 }{ 2 \pi \sigma\s{th} } \exp \left[ - \frac{ \delta x^2 + \delta p^2 }{ 2 \sigma\s{th} } \right].
\end{equation}

Assuming the joint evolution of the mechanical oscillator and bath to be unitary, one can write the WF of the composite system (mechanical oscillator+bath)
\begin{equation}
	W ( x' , p' , \delta x , \delta p )
	= W\s{m} ( x [x',p',\delta x , \delta p ] , p [x',p',\delta x , \delta p ] )
	\times W\s{B} (\delta x , \delta p ).
\end{equation}
Here the first argument of $W\s{m}$, $x [\dots]$, means the solution of~\cref{eq:io:mech:term:x,eq:io:mech:term:p} for $x$ etc.
Since in the contemporary experiments the quality-factors of the mechanical oscillators can exceed $Q = 10^6$ (see e.g.~\cite{masoncontinuous2019,windeycavitybased2019,deliccavity2019}), one can approximate $\sigma \approx 1$, $\epsilon = 0$, $\zeta = \pi$ with accuracy $o(Q^{-1})$.
Moreover, $\exp[ - \damp \tau\s{m} /2] \approx 1$.
Therefore
\begin{gather}
	x \approx x' - \sqrt \frac{ 2 \pi \damp }{ \Omega\s{m} } \delta x \equiv x' - \theta \delta x,
	\quad
	p \approx p' - \theta \delta p,
    \quad
	\theta \equiv \sqrt{ \frac{ 2 \pi \damp }{ \Omega\s{m} }}.
\end{gather}

To obtain the WF of the mechanical oscillator after this evolution, one has to trace out the degrees of freedom of the environment
\begin{equation}
	W' (x',p') = \iint \dd ( \delta x ) \dd ( \delta p ) W\s{m} ( x' - \theta \delta x , p' - \theta \delta p)
	W\s{B} (\delta x , \delta p ; \sigma\s{th}).
\end{equation}
Making a substitution $( u, v) = \theta \cdot ( \delta x , \delta p)$ we arrive to the simple expression
\begin{equation}
	\label{eq:wf:after:bath}
	W' (x', p') =  \iint \dd u \dd v W\s{m} ( x' - u , p' - v ) W'\s{B} (u , v ),
\end{equation}
where $W'\s{B} (u,v) = W\s{B} (u , v , \sigma\s{th} \theta^{2})$.
The~\cref{eq:wf:after:bath} describes a convolution of the initial Wigner function $W\s{m}$ with a WF of a thermal state, whose variance is reduced by the mechanical $Q$-factor.
Due to this rescaling, $W'\s{B}$ is a very narrow Gaussian distribution, with variance much below $1$.
In this manuscript we use primarily the value $\sigma\s{th} \theta^{2} = 4 \pi \frac{ H\s{m}}{ \Omega\s{m}} \equiv H_0 = 0.002$ which, for an oscillator of eigenfrequency $\Omega\s{m} = 2 \pi \times \SI{100}{\kHz}$ and $Q=10^6$ is equivalent to occupation of the environment equal to $n\s{th} \approx 10^{7}$ phonons.
This is the equilibrium occupation of such an oscillator at the temperature of \SI{50}{\K}.

\section{Robustness of the stroboscopic method} 
\label{sec:robustnessofthestroboscopicmethod}

In this section, to prove the feasibility of the stroboscopic method, we evaluate the robustness of our method against the finiteness (non-instantaneousness) of the duration of the nonlinearity and the impact of the thermal noise.
It is convenient to measure the temporal extent $\tau$ of the nonlinear potential application in units of the mechanical motion period $\Theta = \Omega\s{m} \tau$, so that $\Theta = 2 \pi$ corresponds to a full mechanical oscillation.
In Supplementary~\cref{fig:2:robust}~(a) we plot the variance of nonlinear quadrature as a function of the free parameter for different values of the phase rotation $\Theta$.
The plots illustrate that as the phase gain $\Theta$ from the quadratic evolution decreases, the result of the combined dynamics approaches the action of the pure nonlinearity.
In particular, for the value in the figure, difference in the minimal value of the nonlinear variance is less than 10\%.
As the strength of the applied nonlinear potential increases, the nonlinear squeezing becomes more sensitive to the free rotation.
The reason is apparently because the state with stronger nonlinearity is further from the classical domain and hence more fragile to perturbations.

In Supplementary~\cref{fig:2:robust}~(b) we show the nonlinear variance from Supplementary~\cref{fig:2:robust} for the value $\Theta = \pi/20$ after one half of mechanical period in the thermal environment.
We can see that in the setup of the present experiment with levitated nanoparticles~\cite{deliccooling2020}, for which $H = 10^2 H_0$ most of the nonlinear squeezing is lost, however for a narrow region of the parameter values the mechanical state can still be used as a resource of the nonlinear squeezing.
In case of the heating rate being suppressed tenfold, the nonlinear squeezing persists in the thermal environment.
It has to be noted that in electromechanical or optomechanical experiments with bulk oscillators, typical heating rates are much smaller.
For instance, in electromechanical experiment~\cite{palomakientangling2013}, $H\s{m} = 1.4 \times 10^{-3} \Omega\s{m}$, and in the experiment with optomechanical crystal reported in Ref.~\cite{riedingerremote2018}, $H\s{m} < 0.1 \times 10^{-3} \Omega\s{m}$.

\section{On duration of the nonlinear potential application}
\label{sec:ondurationofthenonlinearpotentialapplication}

One possibility to use the nonlinear potential that seems straightforward, is to apply it simultaneously with the quadratic one and investigate the steady state.
This strategy, unfortunately does not allow seeing coherent effects of nonlinearity.
In this section we explore application of nonlinear potential for durations comparable to the mechanical period and show that the stroboscopic application is optimal.

First, we have to note that a mechanical oscillator in a sum of quadratic and an odd-order potentials will have a steady state only in the case of weak cubic nonlinearity when there exists a local minimum.
Then the oscillator is going to be trapped in its vicinity where the effective potential is a displaced quadratic one.
Therefore, we only have to consider finite-time action of the cubic potential and show that increasing the duration of nonlinear potential's action decreases its effect.

The unitary dynamics of an oscillator in presence of both quadratic and polynomial potential is described by the operator which in our notations for the nonlinear gain $\gamma$ can be written as
\begin{equation}
    \oper U (\tau) = \exp\left( - \ii \left[ \frac \Omega 4 ( p^2 + x^2 ) + \frac \gamma {2 k \tau} x^k \right] \tau \right).
\end{equation}
The resulting action of this unitary is determined by two parameters: total phase rotation $\Omega \tau$ and nonlinear gain $\gamma$.
In the experiment, however, the parameters that are defined by the external control (e.g., optical driving) are the stiffnesses of quadratic and polynomial potentials, respectively, $\Omega$ and $\varkappa = \gamma/\tau$.

\begin{figure}[htb!]
    \centering
    \includegraphics[width=.95\linewidth]{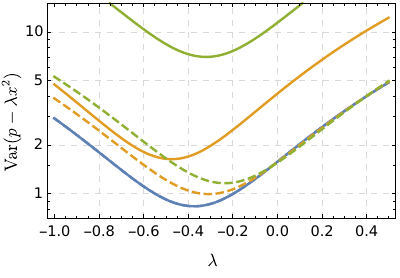}
    \caption{Nonlinear squeezing for increased duration of the nonlinearity following different approaches.  Solid lines, following~\cref{eq:samestiffness}: duration increases preserving the stiffness of nonlinearity.  Dashed lines, following~\cref{eq:samegain}: duration increases keeping the nonlinear gain intact.  Different colors correspond to duration $\tau'$ proportionally increased compared to the value $\tau_0$ from main text.  Blue lines $\tau' = \tau_0$, yellow: $\tau' = 2 \tau_0$, green: $\tau' = 3 \tau_0$.  As $\tau'$ increases further, both methods approach the result of a noisy thermal mechanical state.}%
    \label{fig:steadystate}
\end{figure}

In this light, there are two different ways to compare longer interactions:
\begin{itemize}
    \item Assuming constant stiffnesses but longer duration
        \begin{equation}
            \oper U(\tau') = \exp\left( -i \left[ \frac \Omega 4 ( p^2 + x^2 ) + \frac{ \gamma }{ 2 k \tau } x^k \right] \tau' \right).
            \label{eq:samestiffness}
        \end{equation}
        One could expect improvement in nonclassicality generation from this strategy as longer durations here cause larger gains of nonlinearity.
        Surprisingly, after a certain value of the gain, stronger nonlinearities are cancelled by the quadratures mixing caused by free rotation.
        Moreover, in an experiment, too high gain can cause the partice to escape the trap, or damage to the setup in case of a bulk system.
        Indeed, the original stroboscopic application assumed very short strong nonlinearity that, if applied for significantly longer durations, can exceed reasonable bounds.
    \item Assuming longer interaction but with proportionally reduced nonlinear stiffness, that causes equal nonlinear gain
        \begin{equation}
            \oper U (\tau') = \exp \left( - \ii \left[ \frac \Omega 4 ( p^2 + x^2 ) + \frac{ \gamma }{ 2 k \tau' } x^k \right] \tau' \right).
            \label{eq:samegain}
        \end{equation}
        This way, equal nonlinearity faces larger rotation in the phase space, so no improvement is expected.
\end{itemize}
These findings are summarized at Supplementary~\cref{fig:steadystate} where squeezing of the nonlinear quadrature is plotted against the parameter $\lambda$.
The nonlinear quadrature is computed with respect to the state obtained from the initial squeezed thermal state, same as in the main text, by application one of the unitaries given by~\eqref{eq:samestiffness}~or~\eqref{eq:samegain} assuming cubic nonlinearity ($k = 3$).

Rather expectedly, if the duration of the nonlinearity is increased while preserving the total gain, increase of duration does not increase the nonlinear squeezing.
Eventually the nonclassical effects as negative Wigner function and nonlinear squeezing from nonlinearity become insignificant.

If the nonlinearity of constant stiffness is applied for longer times, one could, in principle, expect stronger nonclassical effects.
Unfortunately, this is not the case, and longer joint application of quadratic and cubic potential results in the mechanical mode being driven to the thermal state with increased occupation.
Importantly, increasing the stiffness of the nonlinearity keeping the duration intact only increases the occupation of the resuting noisy state.
That is, the dephasing action of the free rotation causes the nonlinearity to produce only noise in mechanics even for stiff nonlinear potentials.
In conclusion, our simulations show that the presence of even a weak quadratic term in the potential makes the effect of the cubic nonlinearity vanish entirely.

\section{Expectations of quadrature products~$\avg{\oper x^n \oper p^m}$} 
\label{sec:expectationsofquadratureproducts_q^n p^m_}

In this section we describe a way to compute the expectation of an operator $\oper x^n \oper p^m$.
This material can as well be found in textbooks such as~\cite{schleichquantum2011,gerry_introductory_2004}.

Following Weyl quantization rules~\cite{hallquantum2013}, one can connect the classical expectation $\avg{ x^n y^m }\s{cl}$ with the expectation of a combination of the operators $\oper x^r \oper p^m \oper x^{n - r }$:
\begin{equation}
	\label{eq:weyl}
	2^n \avg{ x^n y^m }\s{cl} = \sum \limits_{r = 0}^n \pmt n r \avg{ \oper x^r \oper p^m \oper x^{n - r } }.
\end{equation}
In this equation the classical and quantum expectations $\avg{\bullet}\s{cl}$ and $\avg{ \bullet }$, respectively, are computed as
\begin{gather}
	\avg{ f (x,y) }\s{cl} = \iint \dd x \dd y W ( x ,y ) f(x,y),
	\\
	\avg{ \oper f } = \Tr[ \rho \oper f ],
\end{gather}
where $W$ is the Wigner function corresponding to the state $\rho$.

\cref{eq:weyl} is not particularly useful when one requires to compute a product of the form $\avg{ \oper x^n \oper p^m}$, e.g., $\avg{ \oper x^2 \oper p }$.
The efficient computation can be done recurrently.

First, note that
\begin{equation}
	\ox^r \op^m \ox^{ n - r } = \ox^n \op^m + \ox^r \comm{ \op^m }{ \ox^{  n - r }},
\end{equation}
therefore
\begin{equation}
	\sum \limits_{r = 0}^n \pmt n r \avg{ \oper x^r \oper p^m \oper x^{n - r } }
	= \sum \limits_{ r = 0 }^{n } \pmt n r \left( \ox^n \op^m + \ox^r \comm{ \op ^m }{ \ox^{ n - r }} \right)
	= 2^n \ox^n \op^m + \sum \limits_{ r = 0 }^{n } \pmt n r \ox^r \comm{ \op ^m }{ \ox^{ n - r }}.
\end{equation}
Importantly, all the terms with commutators in this expression simplify to expressions whose total power is smaller than $n + m $.
This allows to write a recurrent relation
\begin{equation}
	\label{eq:quadraturemoments}
	\avg{ \ox^n \op^m } = \avg{ x^n p^m }\s{cl} - \frac{ 1 }{ 2^n } \sum \limits_{ r = 0 }^{n - 1} \pmt n r \avg{\ox^r \comm{ \op ^m }{ \ox^{ n - r }}}.
\end{equation}
The upper limit of the sum is decreased because for $r = n$, $\comm{ \op^m }{ \ox^{ n - r } } = \comm{ \op^m }{ \II } = 0$.

The equation above connects the expectation of the quantum operator $\ox^n \op^m$ with a classical value $\avg{ x^n p^m }\s{cl}$ and a combination of expectations of operators of lower total power.
Substituting~\cref{eq:quadraturemoments} into the right part of itself one at the end arrives to an expression that consists entirely of classical expectations.

To illustrate this method, let us compute the variance of the nonlinear quadrature $\op - \lambda \ox^2$ in the quantum state~$\rho$, which involves computing the quantum expectation of~$\ox^2 \op$.

First, we write the expression for the variance of the nonlinear quadrature
\begin{multline}
	\Var( \op - \lambda \ox^2 ) = \avg{ \left( \op - \lambda \ox^2 \right)^2 } - \avg{ \op - \lambda \ox^2 }^2
	= \avg{ \op^2 } - \lambda \left(  \avg{ \op \ox^2 } + \avg{ \ox^2 \op } \right) + \lambda^2 \avg{ \ox^4 }
	\\
	- \avg{ \op }^2
	+ 2 \lambda \avg{ \op } \avg{ \ox^2 } - \lambda^2 \avg{ \ox^2 }^2.
\end{multline}
The expectations of the powers of individual quadratures directly map to the classical expectations
\begin{equation}
	\avg{ \op^n } = \avg{ p^n }\s{cl},
	\quad
	\avg{ \ox^n } = \avg{ x^n }\s{cl}.
\end{equation}
For the cross-terms, on the one hand, one can write a simple chain of equations
\begin{equation}
	\avg{ \ox^2 \op + \op \ox^2 } = \frac 12 \left( \ox^2 \op + 2 \ox \op \ox + \op \ox^2 \right) = 2 \avg{ x p^2 }\s{cl}.
\end{equation}

For the purpose of illustrating~\cref{eq:quadraturemoments} we first reorder the cross-term to have the powers of $\ox$ on the left
\begin{equation}
	\label{eq:asymmetrize}
	\ox^2 \op + \op \ox^2 = 2 \ox^2 \op - 2 \ox \comm{ \ox }{ \op }.
\end{equation}
The first term can be evaluated as
\begin{equation}
	\label{eq:expectx2p}
	\avg{ \ox^2 \op } = \avg{ x^2 p }\s{cl}
	-
	\frac 14 \left( \avg{ \comm{\op }{\ox^2 } + 2\ox \comm{\op }{\ox } }
	\right)
	= \avg{ x^2 p }\s{cl} - \frac 14 \left( - 4 \avg{ \ox \comm{\ox}{\op} }  \right)
	= \avg{ x^2 p }\s{cl} + \avg{ \ox } \comm{ \ox }{ \op }.
\end{equation}
Combining~\cref{eq:asymmetrize,eq:expectx2p} yields
\begin{equation}
	\avg{ \ox^2 \op + \op \ox^2 } = 2 \avg{ x p^2 }\s{cl}.
\end{equation}


\newpage
\renewcommand\thesection{\fpeval{\arabic{section}-\presupsections}}
\section*{Supplementary Figures}

\begin{figure}[htb]
	\centering
	\includegraphics[width=.95\linewidth]{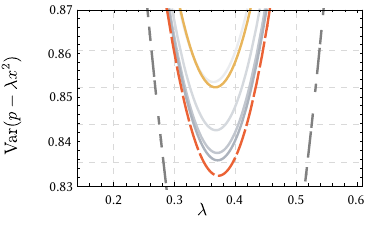}
	\caption{Convergence of the Suzuki-Trotter approximation.
		Gray dot-dashed line represents the ideal unitary nonlinearity.
		Red dashed line comes from simulation in Fock space.
		Yellow is Method 2 with Trotter number $N = 36$.
		Different shades of gray is Method 1 with Trotter number $N = 6,12,24,36$.
		As $N$ increases, the color becomes darker.
	}%
	\label{fig:convergence}
\end{figure}

\begin{figure}[t]
	\centering
	\includegraphics[width=.99\linewidth]{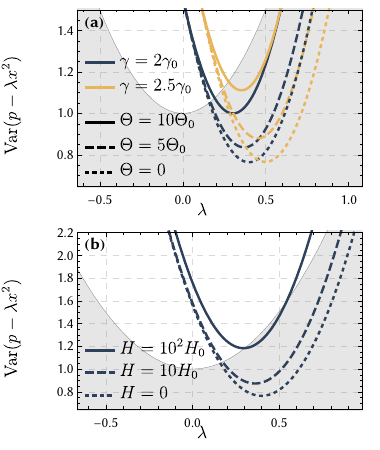}
	\caption{
		Robustness of the nonlinear squeezing against main imperfections for the cubic potential $V(x) \propto \gamma x^3$.
		(a) Robustness against the duration $\Theta$ of quadratic evolution for different values of nonlinear gain $\gamma$.
		The mechanical oscillator is initialized in a squeezed thermal state (mean occupation $n_0 = 0.05$, squeezing $s = 1.2$) and evolves according to~\eqref{eq:suzuki-trotter} where $\Omega\s{m} \tau = \Theta$.
		Parameters same as in~\cref{fig:1:application} $\gamma_0 = 0.2$, $\Theta_0 = \pi /100$.
		(b) Robustness against thermal noise for different values of heating rate $H\s{m}$.
		After the evolution illustrated in~(a), the mechanics is subject to thermal environment for a half of oscillation period.
		The parameters are $H_0 = 0.002$, $\gamma = 2 \gamma_0$, $\Theta = 5 \Theta_0$.
		Thin gray line shows variance computed at vacuum state, values below (filled area) provide advantage over vacuum as a resource for implementation of a measurement-based gate~\cite{marekgeneral2018}.
	}
	\label{fig:2:robust}
\end{figure}

\end{document}